\documentclass[12pt]{amsart}
\usepackage{amsmath}
\usepackage{amssymb}
\usepackage[latin1]{inputenc}
\usepackage[T1]{fontenc}
\newtheorem{lemma}{Lemma}

\newtheorem{remark}{Remark}

\newcommand{\tr}{{\rm tr }}
\newcommand{\lh}{\mathcal{L(H)}}
\renewcommand{\th}{\mathcal{T(H)}}

\newcommand{\ket}{\rangle}
\newcommand{\vp}{\varphi}

\newcommand{\C}{\mathbb{C}}

\newcommand{\N}{\mathbb{N}}

\newcommand{\R}{\mathbb{R}}
\newcommand{\Z}{\mathbb{Z}}
\newcommand{\be}{\begin{equation}}
\newcommand{\eeq}{\end{equation}}
\newcommand{\bet}{\begin{equation*}}
\newcommand{\eeqt}{\end{equation*}}
\newcommand{\bea}{\begin{eqnarray}}
\newcommand{\eeqa}{\end{eqnarray}}
\newcommand{\beat}{\begin{eqnarray*}}
\newcommand{\eeqat}{\end{eqnarray*}}

\newcommand{\h}[1]{\mathcal{#1}}
\newcommand{\hil}{\mathcal{H}}
\newcommand{\hi}{\mathcal{H}}

\newcommand{\hB}{\mathcal{B}}

\newcommand{\cc}[1]{\overline{#1}}

\newcommand{\pa}{\partial}
\newcommand{\la}{\lambda}
\newcommand{\kb}[2]{|#1\rangle\langle#2|} 
\def\<{\langle}
\def\>{\rangle}

\begin{document}

\title{State reconstruction formulas for the $s$-distributions and quadratures}

\author{Jukka Kiukas}
\address{Institute for Theoretical Physics, University of Hannover, Hannover, Germany}
\email{jukka.kiukas@utu.fi}
\author{Juha-Pekka Pellonp\"a\"a}
\address{Department of Physics and Astronomy, University of Turku, Turku, Finland}
\email{juha-pekka.pellonpaa@utu.fi}
\author{Jussi Schultz}
\address{Department of Physics and Astronomy, University of Turku, Turku, Finland}
\email{jussi.schultz@utu.fi}
\begin{abstract}
We consider the method of infinite matrix inversion in the context of quantum state reconstruction. Using this method we give rigorous proofs for reconstruction formulas for the Cahill-Glauber $s$-parametrized distributions and the rotated quadrature distributions. We also demonstrate how to construct the $s$-distributions from the quadrature data.

\end{abstract}
\maketitle

\section{Introduction}
Given a quantum system in a state $\rho$, that is, a positive operator of trace one acting on a Hilbert space $\hil$, one can perform measurements on the system to obtain
probability measures corresponding to various observables being measured.
The inverse problem, namely, the reconstruction of an unknown state from some set of measured probability distributions is one of the important problems in quantum theory, and consequently, it has been studied extensively. Here we are only interested in the case where the Hilbert space is infinite dimensional; this is typical in quantum optics (for an overview, see e.g. \cite{Welsch} or \cite{Leonhardt} and references therein). The ''inversion'' character of the problem can be formulated in different mathematical forms. Probably the most commonly used approach is to use some integral transform to convert the measured distributions into a desired quasiprobability distribution. The usual tomographic scheme uses the inverse Radon transform to reconstruct the Wigner function from the experimentally obtainable rotated quadrature distributions \cite{VR1989, Bertrand1987, Smithey1993}.
In the context of cavity QED and ion trapping, an alternative approach using an integral transform of the Rabi oscillations of a two-level atom coupled to the field has also been proposed \cite{Leibfried1996,Lougovski2003}. Even though the quasiprobability distributions, such as the Wigner function, contain complete information about the quantum state of the system, and can be used to calculate expectation values of observables, it is of interest to also reconstruct directly the actual density operator of the system \cite{Dariano1994, DAriano, Leonhardt1996, Kuhn1994, Mancini1997, DeNicola, Paris}. In \cite{DAriano}, an explicit reconstruction formula for the quadrature distributions was given. This was later generalized to cover the distributions of generic linearly transformed quadratures \cite{DaMa1996}. The recent progress in the field of quantum state reconstruction has been reviewed in \cite{Welsch}.

The purpose of this paper is to illustrate the use of the method of infinite matrix inversion in reconstructing the density matrix of a quantum system from certain, measured or otherwise obtained, phase space distributions. This means that we are using a fixed computational basis of the (infinite dimensional) Hilbert space in question, and have to deal with various convergence issues related to the matrix inversion. These will then naturally lead to conditions on the state under which the formulas are valid.

The structure of the paper is the following.
In section \ref{infotayd} we fix the notations, describe the basic idea in the derivation of the reconstruction formulas, and discuss some general issues associated with state reconstruction, infinite matrices, and approximating them by finite matrices.
Section \ref{kvadratuuri} is devoted to state reconstruction from the rotated quadrature distributions; we will give an alternative, mathematically rigorous derivation for the formula appearing in \cite{DAriano}, and also demonstrate how only a finite number of quadratures need to be measured in the case where the density matrix is known to be finite. In section \ref{lambda} we consider the Cahill-Glauber $s$-distributions. First we show how they can be constructed from quadrature data, and then we consider two methods for obtaining reconstruction formulas.
 Since the $s$-distributions are not in general positive, they are not strictly measurable quantities in the sense of the theory of measurement. However, they can indirectly be constructed from actual measurements. We return to this question in section \ref{lambdadistr}.


\section{Preliminaries on state reconstruction and infinite matrices}\label{infotayd}

Let $\hil$ be a complex infinite-dimensional Hilbert space; we fix an orthonormal basis
$\{|n\ket\mid n\in \N\}$ of $\hil$. (Here we denote $\N := \{0,1,2,\ldots\}$.) 
This computational basis is identified with the photon number basis, or Fock basis, in the case where $\hil$ is associated with a single mode
electromagnetic field. 
We will, without explicit indication, use the coordinate representation, in which $\hil$ is represented as $L^2(\R)$ via the unitary map $\hil\ni\vert n\rangle\mapsto h_n\in L^2(\R)$, where $h_n$ is the $n$th Hermite function.

Let $a$ and $a^*$ denote the raising and lowering operators associated with the above
basis of $\hil$, and
define the operators $Q:=\frac{1}{\sqrt{2}}(a^*+a)$ and
$P:=\frac{i}{\sqrt{2}}(a^*-a)$ which, in the coordinate
representation, are the usual multiplication and differentiation operators, respectively:
$(Q\psi)(x) = x\psi(x)$ and $(P\psi)(x)= -i\frac{d\psi}{dx}(x)$.\footnote{To be precise, one has to take the operator closure when defining $Q$ and $P$ in terms of $a$ and $a^*$, but this is a well-known technical issue which is not important here.}
In the case of the electromagnetic field, $Q$ and $P$ are called the
quadrature amplitude operators of the field. 
The selfadjoint operator $N:=a^*a$ is the (photon) number operator; it generates 
the phase shifting unitaries $R(\theta):=e^{i\theta N}$, $\theta\in [0,2\pi)$, and we can define the
rotated quadrature operators $Q_\theta$ by
$$
Q_\theta = R(\theta)QR(\theta)^*, \ \theta\in [0,2\pi).
$$
We will also need the displacement operator
$D(z)=e^{z a^*-\overline z a}$, $z\in\C$, for which holds $D(z)^*=D(z)^{-1}=D(-z)$ and
$R(\theta)D(z)R(\theta)^*=D\big(ze^{i\theta}\big)$. The latter implies $D(re^{i\theta}) =R(\theta)D(r)R(\theta)^\ast$, which will be important. 

Let $\lh $ be the set of bounded
operators on $\hil$, and $\th $ the set of trace class operators. The former space is equipped with the operator norm $\|\cdot\|$, and the latter with trace norm $\|\cdot\|_1$. Since $\hil$ is associated with a
quantum system, these sets have physical meaning: the states of the system are represented by positive operators $\rho\in \th $ with the unit trace. The state operator is uniquely determined by the numbers $\rho_{nm}:=\langle n|\rho |m\rangle$, constituting the density matrix in the number basis. The pure states correspond to projections onto the one-dimensional subspaces of $\hil$, and are thus of the form $\vert\vp\rangle\langle\vp\vert$, where $\vp\in\hil$ is a unit vector. In particular, we will need the coherent states $|z\rangle$, defined for each $z\in\C$ by
$$
\vert z\rangle :=D(z)\vert 0\rangle =e^{-\frac{|z|^2}{2}} \sum_{n=0}^\infty \frac{z^n}{\sqrt{n!}} \vert n\rangle.
$$

When reconstructing an unknown state of the system directly from some measurement data, one of course requires that the data determines the state uniquely. The set of measured observables is then said to be \emph{informationally complete} \cite{Prugovecki,Busch}.
The observables in quantum mechanics are represented by normalized positive operator measures (POMs) $\mathsf{E}$
defined on a $\sigma$-algebra $\Sigma$ of subsets of some outcome set $\Omega$.\footnote{Normalized positive
operator measure (POM) is a map $\mathsf{E}:\,\Sigma\to \lh $ 
which is $\sigma$-additive in the weak operator topology, and has the property $\mathsf{E}(\Omega)=I$ (the identity operator).} The probability measure $X\mapsto\tr[\rho \mathsf{E}(X)]$ corresponding to a state $\rho $ and a POM $\mathsf{E}$ is then the one according to which the measurement outcomes are distributed.  In this article, $\Omega$ is always a subset of either $\R$ or $\R^2\cong\C$, and $\Sigma$ is the associated Borel $\sigma$-algebra $\h B(\Omega)$.
The most common observables are of the conventional von Neumann type,
that is, normalized projection valued measures or, in the case when $\Omega=\R$, selfadjoint
operators in $\hil$. In particular, the spectral measure of $Q_\theta$ is such an observable; we will denote it by $\mathsf{Q}_{\theta}:\h B(\R)\to \lh $.


We consider the reconstruction of the state $\rho  $ by determining the elements of the density matrix $(\rho_{nm})$. Before proceeding to the derivation of the reconstruction formulas, we will make some general remarks on state reconstruction in infinite-dimensional spaces.

A typical way of dealing with infinite matrices is to approximate them by finite matrices. In the case of a density operator $\rho$, a natural measure of approximation is the trace norm: for the projection $P_p:=\sum_{n=0}^{p-1}\kb n n$, $p=1,2,3,\ldots $, we can define
$$
\rho _p:=\frac1{\tr(P_p\rho P_p)}P_p\rho P_p
$$ 
for all states $\rho \in\th$ such that $\tr(P_p\rho P_p)=\sum_{n=0}^{p-1}\rho _{nn}\ne0$. Note that for any $\rho \in\th$,
 there exists a smallest $p_0\in \{1,2,\ldots\}$ such that $\tr(P_p\rho P_p)\ne 0$ when $p\ge p_0$. The following well-known lemma implies that we can approximate a state $\rho $ in the trace norm by a state $\rho_p$, which has only finitely many non-zero matrix elements:
 \begin{lemma} Let $\hi$ be a separable Hilbert space, and $(P_n)_{n\in \N}$ an increasing sequence of projections on $\hi$, converging strongly to the identity operator. Then for each trace class operator $\rho $ on $\hi$, the sequences $(P_n\rho )_{n\in\N}$, $(\rho P_n)_{n\in\N}$, and $(P_n\rho P_n)_{n\in\N}$ converge to $\rho $ in the trace norm.
\end{lemma}

One can also use the trace of $\rho $ to quantify how well a truncation $\rho_p$ approximates $\rho $: Suppose that we have determined the first $p$ diagonal elements $\rho_{nn}$, $n=0,1,...,p-1$, and that $\sum_{n=0}^{p-1}\rho_{nn}>1-\epsilon$. Then, since $\tr\;\rho=1$ one must have $\rho_{nn}<\epsilon$ for all $n\ge p$. The positivity of $\rho $ implies $|\rho_{mn}|^2\le \rho_{mm}\rho_{nn}$ for all $n,\;m\in\N$, so $|\rho_{mn}|<\epsilon$ for all $n\ge p$ or $m\ge p$.
Obviously, this has a simple physical meaning in the case of an electromagnetic field: the condition $\sum_{n=p}^\infty \rho_{nn}<\epsilon$ just says that the probability of measuring a photon number larger than $p$ in the state $\rho$ is less than $\epsilon$. In many concrete applications, one can choose a finite $p\in \N$ such that it is practically impossible ($\epsilon\approx 0$) to get a photon number larger than $p$; in any case, no measuring apparatus can detect arbitrary high energies. This then implies that $\rho_{mn}=0$ for all $n\ge p$ or $m\ge p$, suggesting that we may \textit{a priori} assume that the state matrix is finite.

However, typically the maximal detectable photon number is not actually known, or depends on the construction of the measuring apparatus; hence fixing it to a finite value represents an artificial truncation of a system which is genuinely infinite-dimensional. In view of the state reconstruction, this is particularly significant, since the state is not known beforehand. There is no way of knowing how large number $p$ one must take so as to get $\sum_{n=p}^\infty \rho_{nn}<\epsilon$ for a given $\epsilon$. 

 Next we describe how to reduce the reconstruction problem into the infinite matrix inversion in the case of some phase space quasi-probability distributions. The distributions we are considering in this paper have densities of the form
$$
W_\rho(r,\theta)=\sum_{m,n=0}^\infty \rho_{mn}e^{i(n-m)\theta}f_{nm}(r)
$$ 
where the  $f_{nm}$ are some functions of the radial parameter $r$, and $\theta$ is the angle coordinate. The general heuristic idea to obtain reconstruction formulas from such distributions is the following. Integrating against exponentials $e^{ik\theta}$, for a fixed $k\in \N$, with respect to $\theta$ over $[0,2\pi)$, one is left with a single sum
\begin{equation}\label{summa}
W_{\rho,k}(r):=\frac{1}{2\pi}\int_0^{2\pi}e^{ik\theta}W_\rho(r,\theta)d\theta = \frac{1}{2\pi}  \sum_{m,n=0}^\infty \rho_{mn} \int_0^{2\pi} e^{i(k+n-m)\theta}f_{nm}(r)\, d\theta= \sum_{n=0}^\infty  \rho_{n+k,n}f_{n,n+k}(r).
\end{equation}
We consider two methods of converting this to a infinite matrix equation: integrating \eqref{summa} with respect to some functions $g_{l,k}$ of $r$ to get
\begin{equation}\label{integration}
\int g_{l,k}(r) W_{\rho,k}(r) \, dr = \sum_{n=0}^\infty  \rho_{n+k,n}\int g_{l,k}(r) f_{n,n+k}(r)\, dr,
\end{equation}
or differentiating $l$ times with respect to $r$ at $r=r_0$, which gives
\begin{equation}\label{differentiation}
\frac{d^l}{dr^l}[g_k(r)W_{\rho,k}(r)]|_{r=r_0} = \sum_{n=0}^\infty  \rho_{n+k,n} \frac{d^l}{dr^l} [g_k(r)f_{n,n+k}(r)]|_{r=r_0},
\end{equation}
where $g_k$ are some functions. With fixed $k$, both of these are now infinite matrix relations of the form
\begin{equation}\label{matrixrelation}
y_l =  \sum_{n=0}^\infty a_{ln} x_n,
\end{equation}
and the idea is to choose the functions $g_{k,l}$ or $g_k$ in such a way that this relation can be inverted as
\begin{equation}\label{inversematrixrelation}
x_n=\sum_{l=0}^\infty b_{nl} y_l.
\end{equation}
Of course, one has to take care of the convergence of the series, and other technical details; this has to be done separately in each case. In particular, even if $(a_{ln})$ has a \emph{formal inverse} $(b_{nm})$, i.e. the relation
$$
\sum_{n=0}^\infty a_{ln}b_{nm} = \sum_{n=0}^\infty b_{ln}a_{nm} = \delta_{lm}
$$
holds, it is still not clear if \eqref{inversematrixrelation} is true. In order to briefly demonstrate pathological situations that could in principle arise, we will give the following example at this point: Take
\begin{align*}
(a_{nm})& =
\begin{pmatrix}
1 & 1 & 0 & 0 & 0 &\cdots \\
0 & 1 & 1 & 0 & 0 &\cdots \\
0 & 0 & 1 & 1 & 0 &\cdots \\
0 & 0 & 0 & 1 & 1 &\cdots \\
0 & 0 & 0 & 0 & 1 &\cdots \\
\vdots & \vdots & \vdots & \vdots & \vdots & \ddots \\
\end{pmatrix}, & 
(b_{nm}) &=
\begin{pmatrix}
1 & -1 & 1 & -1 & 1 &\cdots \\
0 & 1 & -1 & 1 & -1 &\cdots \\
0 & 0 & 1 & -1 & 1 &\cdots \\
0 & 0 & 0 & 1 & -1 &\cdots \\
0 & 0 & 0 & 0 & 1 &\cdots \\
\vdots & \vdots & \vdots & \vdots & \vdots & \ddots \\
\end{pmatrix}.
\end{align*}
Clearly, these matrices are formal inverses of each other, and the relation \eqref{matrixrelation} is well-defined for any complex sequence $(x_n)$. The sequence $(y_l)$ is then given by $y_l = x_l+x_{l+1}$. Consider the following two cases:
\begin{itemize}
\item[(a)] Let $x_n=1$, $n\in \N$. Then $y_n=2$, $n\in \N$, and the relation \eqref{inversematrixrelation}
does not make any sense, since the associated series does not converge for any $n$ (its partial sums form the sequence $2,0,2,0,2,\ldots$).
\item[(b)] Let $x_n=(-1)^n$, $n\in \N$. Then $y_n=0$, $n\in \N$, so the series in \eqref{inversematrixrelation} is well-defined, converging to $0$ for any $n$. But this is not equal to $x_n$,
so \eqref{inversematrixrelation} does not hold.
\end{itemize}
It is easy to see that in this simple example \eqref{inversematrixrelation} holds exactly when $\lim_{n\rightarrow\infty} x_n=0$. For a detailed treatment of infinite matrices, we refer the reader to \cite{Cooke}.

\section{Balanced homodyne detection and quadrature distributions}\label{kvadratuuri}
We begin with a short review of the precise mathematical formulation of the balanced homodyne detection, which is a well-known scheme of measuring quadratures, and is used quite frequently in quantum optics (see e.g. \cite{Leonhardt}).

The description of the homodyne measurement is not entirely straightforward:
for any interval $\h I\subset \R$, the probability $\tr[\rho  \mathsf{Q}_{\theta}(\h I)]$ for the quadrature measurement can be obtained only as a \emph{limit} of balanced homodyne detection measurements.
The reader should consult \cite{Leonhardt} for the basic description; we also wish to mention a mathematically rigorous proof given in \cite{KL2008}.

Balanced homodyne detector consists of a beam splitter, with a pair of photon counters
at the output ports. The signal light beam, and a coherent auxiliary beam
are sent into the beam splitter, and the suitably scaled difference
between the photon numbers at the output ports is recorded as an outcome of
the measurement. The scale factor is the amplitude of the auxiliary beam.

The idea in the mathematical description of \cite{KL2008} is
the following. One considers a sequence of measurements, each with larger
auxiliary field amplitude than the preceding one and such that
the amplitudes grow without bound. From the data of each
measurement, one calculates the probability that the outcome lies in $\h I$.
The sequence of the probabilities
thus generated will converge to $\tr[\rho  \mathsf{Q}_{\theta}(\h I)]$
in the limit where the amplitude tends to infinity regardless of the state $\rho$. Hence, for any fixed state, and any fixed interval, the correct probability can be obtained with arbitrary high precision by using a sufficiently large amplitude.

For each state $\rho$, and each fixed $\theta$, the probability measure $X\mapsto \tr[\rho \mathsf{Q}_\theta (X)]$ has a density function
$x\mapsto W_{\rho}^{\textsf{qd}}(x,\theta)$. It is easy to see that the function $W_\rho^{\textsf{qd}}: \R \times [0,2\pi)\to \C$ so defined is actually measurable, and integrable\footnote{That $W_\rho^{\textsf{qd}}$ may be chosen to be measurable can be seen by writing the density operator in the spectral representation and using the fact that $\theta\mapsto R(\theta)$ is weakly continuous. Integrability follows from Fubini's theorem.}. Now we can write
\begin{equation}\label{quadraturerelation}
\int g(x) W_{\rho,k}^{\textsf{qd}}(x)\, dx = \sum_{n=0}^\infty \rho_{n+k,n}  \int g(x) f_{n,n+k}(x)\, dx,
\end{equation}
where
\begin{eqnarray*}
W_{\rho,k}^{\textsf{qd}}(x) &:=& \frac{1}{2\pi}\int_0^{2\pi} e^{ik\theta}W_\rho^{\textsf{qd}}(x,\theta)\, d\theta;\\
f_{nm}(x) &:=& h_n(x)h_m(x),
\end{eqnarray*}
and $g:\R\to \C$ is any bounded measurable function. The technical details leading to \eqref{quadraturerelation} are straightforward (see e.g. \cite[Lemma 5]{KLP2008}). The relation \eqref{quadraturerelation} now corresponds to \eqref{integration}, once we have chosen suitable functions $g=g_{l,k}$.

By doing this, we will end up with the known reconstruction formula (see \eqref{reconstruction} below) which, up to our knowledge, was first given by Leonhardt and D'Ariano \cite{DAriano}, who wanted to replace the traditional and mathematically troublesome inverse Radon transform scheme with a direct reconstruction of the density matrix in terms of the quadrature data. The same formula was later obtained as a special case of more general group theoretical results by Cassinelli {\em et al}. \cite{Cassinelli}

Instead of using their methods, we derive the formula by using matrix inverse relations. In our opinion, this method more explicitly illustrates the technique by which the matrix elements are picked out by the suitable averaging functions, namely the derivatives of the so called \emph{Dawson's integral}. Most of the construction is already given in our recent paper \cite{KLP2008}, where we gave a direct proof for the fact that the set of quadrature observables is informationally complete. Using this method, we will also explicitly demonstrate the fact that if the state matrix is \textit{a priori} assumed to contain only a finite number of nonzero elements, then it is only required to measure a finite number of quadratures.

Now consider the function $Y:\R\to \R$, defined simply by
$Y(x)=2\frac{d}{dx}\,{\rm daw}(x)$, where
$${\rm daw}(x)=e^{-x^2}\int_0^x e^{t^2}\, dt$$
is the well-known \emph{Dawson's integral} (see e.g. \cite[pp. 298-299]{Abramowitz} or \cite[Chapter 42]{Atlas}).
We are interested in the derivatives $Y^{(p)}$, $p\in\N$. Since
they appear in many physical situations, such as in a series expansion of the so called Voigt function
in spectroscopy (see e.g. \cite[p. 69-70]{Armstrong}), the
computational evaluation of $Y^{(p)}$
has been studied extensively (see e.g. \cite{Milone, Armstrong, McCabe,Schreier}).

According to Lemma 1 of \cite{KLP2008}, each function $Y^{(p)}$ is bounded. Hence we can use the functions $g_{l,k} = Y^{(k+2l)}$. Since each $W_{\rho,k}^{\textsf{qd}}$ is integrable, the integrals
$$
W_{\rho,k,l}^{\textsf{qd}}:=\int_{\R} g_{l,k}(x)W_{\rho,k}^{\textsf{qd}}(x)\, dx
$$
are well-defined.
Putting
$$
c_{ln}(k):= \int_\R g_{l,k}(x)h_{n}(x)h_{n+k}(x)\, dx = \langle n\vert Y^{(k+2l)}(Q) \vert n+k\rangle,
$$
we thus have the matrix relation
\begin{equation}\label{reconstruction0}
W_{\rho,k,l}^{\textsf{qd}} =\sum_{n=0}^l c_{ln}(k) \rho_{n+k,n},
\end{equation}
where the infinite sum is now reduced to a finite one because of the special properties of the Dawson's integral \cite[Lemma 4]{KLP2008}. The coefficients $c_{ln}(k)$ can be evaluated analytically (see Appendix B for the calculation);
the result is
\begin{equation}\label{matrixelements}
c_{ln}(k)=(-1)^{l+n+k}2^{k/2+l}(k+l)!\sqrt{\frac{n!}{(n+k)!}}\binom{l}{n}.
\end{equation}
It is already clear that the elements $\rho_{n+k,n}$ can be solved recursively from \eqref{reconstruction0}. In order to get an explicit formula, we notice that for any fixed $k\in \N$, the matrix relation \eqref{reconstruction0} now assumes the
form
\begin{equation}\label{relation}
y_l = \sum_{n=0}^l (-1)^n\binom ln x_n,
\end{equation}
where
\begin{align}
y_l &= (-1)^{l}\frac{W_{\rho,k,l}^{\textsf{qd}}}{2^l(k+l)!}\,, & x_n &= (-1)^k\sqrt{\frac{2^k n!}{(k+n)!}}\, \rho_{n+k,n}.
\end{align}
According to \cite[p. 43]{Riordan}, the relation \eqref{relation}
can be inverted to give
\begin{equation*}
x_n = \sum_{l=0}^n (-1)^l\binom nl y_l,
\end{equation*}
which immediately yields an explicit reconstruction formula for the density
matrix elements, in terms of the quantities $W_{\rho,k,l}^{\textsf{qd}}$:
\begin{equation}\label{reconstruction}
\rho_{n+k,n}= (-1)^k\sqrt{\frac{(n+k)!}{2^k n!}}\sum_{l=0}^n \binom{n}{l}\frac{W_{\rho,k,l}^{\textsf{qd}}}{2^l(k+l)!},
\ \ n,k\in \N.
\end{equation}
This is the same as formula (34) of Leonhardt {\em et al}. \cite{DAriano}.

\

\noindent\textbf{The finite case.} The above reconstruction formula holds for all states $\rho$, so it is not necessary to assume it finite, in particular. However, when the state matrix is finite, the quadrature data required to determine it is naturally smaller; we close this section by considering the case of finite matrices.

Let $z\in\C$ and $p$ be a positive integer.
Since
$$
\sum_{t=0}^{p-1}z^t=
\begin{cases}
\frac{z^p-1}{z-1}, & z\ne 1, \\
p, & z=1,\\
\end{cases}
$$
if follows that
$$
\sum_{t=0}^{p-1}z^t=0
$$
when $z^p=1$, $z\ne 1$, that is, when $z=e^{i2\pi  q/p}$, $q\in\Z\setminus p\Z$.
For any $t\in\Z$, define $\theta_p^t:=2\pi t/p$. Now the points $e^{i\theta^t_p}$, $t=0,1,...,p-1$,
divide the circle $\mathbb T$ into $p$ parts and 
$$
\frac1p\sum_{t=0}^{p-1}e^{iq\theta^t_p}=
\begin{cases}
0, & q\in \Z\setminus p\Z, \\
1, & q\in p\Z. \\
\end{cases}
$$

Fix $p$ and define for each $k\in\N$ and $X\in\h B(\R)$ an operator $\mathsf{V}^k_p (X)$ by
$$
\mathsf{V}^k_p(X):=\frac{1}{p}\sum_{t=0}^{p-1}e^{ik\theta_p^t}\mathsf{Q}_{\theta_p^t}(X),
$$
so that for a given state operator $\rho$, the experimentally obtainable density
$x\mapsto \tilde{W}^{\textsf{qd},p}_{\rho,k}(x)$ of the measure $X\mapsto \tr[\rho \mathsf{V}^k_p(X)]$ is of the form
$$
\tilde{W}^{\textsf{qd},p}_{\rho,k}(x)= \frac{1}{p}\sum_{t=0}^{p-1}e^{ik\theta_p^t} W_\rho^{\textsf{qd},p}(x,\theta_p^t),
$$
where $W^{\textsf{qd},p}_\rho$ is as before, so it involves only $p$ different quadratures. Now
$$
\langle n|\mathsf{V}^k_p(X)|m\rangle=\langle n|\mathsf{Q}(X)|m\rangle
\frac{1}{p}\sum_{t=0}^{p-1}e^{i(n-m+k)\theta_p^t}, \ \ X\in \h B(\R),
$$
and, hence, by defining $s=n-m+k$,
$$
\langle n|\mathsf{V}^k_p(X)|m\rangle=
\begin{cases}
\langle n|\mathsf{Q}(X)|m\rangle, & m=n+k \mod s, \\
0, & \text{otherwise.}\\
\end{cases}
$$

Suppose now that $\rho_{mn}=0$ when $n,\,m\ge p$. Then
for any $k\in\{0,...,p-1\}$,
$$
\tr\big[\rho \mathsf{V}^k_p(X)\big]=
\sum_{m,n=0}^{p-1}\rho_{mn}\langle n|\mathsf{V}^k_p(X)|m\rangle=
\sum_{n=0}^{p-1-k}\rho_{n+k,n}\langle n|\mathsf{Q}(X)|n+k\rangle.
$$
Using the density $\tilde{W}^{\textsf{qd},p}_{\rho,k}$, we get for any bounded measurable function $g:\R\to\C$
$$
\int g(x) \tilde{W}^{\textsf{qd},p}_{\rho,k}(x)\, dx = \sum_{n=0}^{p-1-k}\rho_{n+k,n} \int g(x) h_{n}(x)h_{n+k}(x)\, dx.
$$
This is in the same form as \eqref{quadraturerelation}, except that the sum is already finite.
Choosing again $g=Y^{(k+2l)}$, $l\in\{0,\ldots, p-1-k\}$, we thus get
\begin{equation}\label{reconstruction11}
\tilde{W}^{\textsf{qd},p}_{\rho,k,l} =\sum_{n=0}^l c_{ln}(k)\rho_{n+k,n},
\end{equation}
where
$$
\tilde{W}^{\textsf{qd},p}_{\rho,k,l}:= \int_\R Y^{(k+2l)}(x)\tilde{W}^{\textsf{qd},p}_{\rho,k}(x)\, dx.
$$
Comparing this with \eqref{reconstruction0}, we evidently get the reconstruction formula
\begin{equation}\label{reconstruction22}
\rho_{n+k,n}= (-1)^k\sqrt{\frac{(n+k)!}{2^k n!}}\sum_{l=0}^n \binom{n}{l}\frac{\tilde{W}^{\textsf{qd},p}_{\rho,k,l}}{2^l(k+l)!},
\end{equation}
where $k\in \{0,\ldots,p-1\}$, $n\in\{0,1,\ldots,p-1-k\}$.
Note that the quantities $\tilde{W}^{\textsf{qd},p}_{\rho,k,l}$ only involve information from the quadratures $P^{Q_\theta}$ corresponding to $\theta\in \{\theta^1_p, \ldots \theta_p^{p-1}\}$.

\section{Cahill-Glauber $s$-parametrized distributions}\label{lambda}
The $s$-parametrized quasiprobability distributions were introduced in quantum optics as mathematical tools \cite{Cahill1969, Glauber1969}, but they have since become accessible also to direct measurements. A scheme for indirect determination of these distributions was suggested by Vogel and Risken \cite{VR1989}, and the pioneering experimental work was done by Smithey {\em et al}. \cite{Smithey1993}.  The $s$-distributions also arise when considering realistic measurements, where the detectors are not assumed to be ideal \cite{Leonhardt1993, Macchiavello1995}. In that case, the measurement outcome statistics correspond to certain $s$-distributions, where the parameter $s$ is related to the efficiency $\eta$ of the detectors, by $s=1-2/\eta$. For our purposes it is convenient to define a parameter $\la$ as $\la=\frac{s+1}{s-1}$. This gives a bijective mapping on $\C\setminus\{1\}$. We will use $\la$ as the parameter for these distributions throughout the paper, and for this reason we will call the $s$-parametrized quasiprobability distributions $\la$-distributions. 

Let $\la\in\C$, $|\la|\le 1$. Define a bounded operator $K^\la:=(1-\la)\sum_{k=0}^\infty\la^k\kb k k$; it has norm $\|K^\la\|=\vert 1-\la\vert$.
If $|\la|<1$ then $K^\la$ is a trace class operator with
$$
\tr[K^\la]=(1-\la)\sum_{k=0}^\infty \la^k=1.
$$
For each $\lambda\in \C$, $|\lambda|\leq 1$, define a (weakly continuous bounded) function $W^\la:\, [0,\infty ) \times [0,2\pi)\to \lh$ by
$W^\lambda (r,\theta)=D(re^{i\theta})K^\la D(re^{i\theta})^*$, where $D(re^{i\theta})$ is the displacement operator. For each state $\rho$ define the $\la$-distribution as the phase space distribution $W^\la_\rho:\, [0,\infty )\times [0, 2\pi) \to \C$,
$$
W^\la_\rho (r,\theta) :=\tr [\rho W^\la (r,\theta) ] = (1-\la)\sum_{k=0}^\infty \la^k \langle k\vert D(re^{i\theta})^\ast \rho D(re^{i\theta}) \vert k\rangle.
$$
Note that $\la=0$ gives us the $Q$-function, and $\la=-1$ gives us the Wigner function (up to a constant scaling factor) of the state.

If $\la\geq 0$, then the operator density $W^\la$ defines a covariant phase space observable 
$$
\mathcal{B}(\C)\ni Z\mapsto \mathsf{G}^{K^\la} (Z) :=\frac{1}{\pi} \int_Z W^\la (r,\theta) r\, drd\theta\in\lh,
$$
in which case the $\la$-distribution can be measured via eight-port homodyne detection. The eight-port homodyne detector consists of two pairs of photon detectors, and the amplitude-scaled photon differences $D_1$ and $D_2$ for each pair are recorded. See \cite{Leonhardt} for the description of the setup measuring the $Q$-function, and note that in our case one has to use the state $K^\la$ instead of the vacuum input in one of the ports (we refer to \cite{eightport} for a detailed description). For other values of $\la$, the $\la$-distribution is not a measurable quantity in the sense of the theory of measurement. However, we will later demonstrate that it can be constructed from the rotated quadrature distributions, which are obtainable via balanced homodyne detection.

Let us consider now the operator valued function $W^\la$. Using the fact that $D(re^{i\theta}) =R(\theta)D(r)R(\theta)^\ast$,  a direct calculation gives us the matrix elements of $W^\la(r,\theta)$ with respect to the number basis:
$$
\langle n \vert W^\la (r,\theta) \vert m\rangle = e^{i(n-m)\theta} K^\la_{nm}(r,\theta),
$$
where 
$$
K^\la_{nm} (r) :=\langle n\vert D(r)K^\la D(r)^\ast \vert m\rangle.
$$
By the formula of Cahill and Glauber \cite{Cahill1969} (for a detailed proof of the formula, see Appendix B), we get
\begin{equation}\label{eq5}
K^\la_{nm}(r)=\sqrt{\frac{n!}{m!}}(1-\la)^{m-n+1}e^{-(1-\la)r^2}r^{m-n}\la^n L^{m-n}_n\big(
(2-\la-\la^{-1})r^2\big)
\end{equation}
where 
$$
L^\alpha_n(x):=\sum_{u=0}^n\frac{(-1)^u}{u!}{{n+\alpha}\choose{n-u}}x^u
$$
is the associated Laguerre polynomial.
Note that the function
\begin{equation}\label{eq6}
\la^n L^{m-n}_n\big((2-\la-\la^{-1})r^2\big)=
\la^n L^{m-n}_n\big(-(1-\la)^2r^2/\la\big)=
\sum_{u=0}^n\frac{\la^{n-u}}{u!}{{m}\choose{n-u}}(1-\la)^{2u}r^{2u}
\end{equation}
can also be defined at $\la=0$ and the extension is smooth with respect to $r$ and $\la$.

Since $\Vert W^\la(r,\theta)\Vert \leq \Vert K^\la \Vert =\vert 1-\la\vert$ for all $r\in [0,\infty)$, $\theta\in [0,2\pi)$, and the mapping $\theta\mapsto W^\la (r,\theta )$ is weakly continuous, we can define the sesquilinear form 
$$
(\psi,\vp)\mapsto \frac{1}{2\pi} \int_0^{2\pi} e^{ik\theta} \langle\psi\vert W^\la (r,\theta) \vp\rangle\,d\theta,
$$
which is clearly bounded (with norm at most $\vert 1-\la\vert$). Thus, for each $k\in\N$, the operator
$$
W^\la_k(r) := \frac{1}{2\pi}\int_0^{2\pi} e^{ik\theta} W^\la (r,\theta)\,d\theta
$$
is well defined as a weak integral. In addition, we have 
$$
W^\la_{\rho,k} (r) := \tr [\rho W^\la_k(r)] =\frac{1}{2\pi} \int_0^{2\pi} e^{ik\theta}W^\la_\rho (r,\theta )\,d\theta,
$$
for each state $\rho$. Since
$$
W^\la_\rho(r,\theta) =\tr[\rho W^\la(r,\theta) ] =\sum_{m,n=0}^\infty \rho_{mn} \langle n\vert W^\la(r,\theta)\vert m\rangle
=\sum_{m,n=0}^\infty \rho_{mn} e^{i(n-m)\theta} K^\la_{nm}(r),
$$
we have
\begin{equation}\label{jakauma}
W^\la_{\rho,k}(r) =\frac{1}{2\pi} \sum_{m,n=0}^\infty \rho_{mn} \int_0^{2\pi} e^{i(k+n-m)\theta} K^\la_{nm}(r)\,d\theta =\sum_{n=0}^\infty \rho_{n+k,n} K^\la_{n,n+k}(r).
\end{equation}
This corresponds to equation \eqref{summa}, and will be the starting point for the reconstruction of the state (see sections \ref{integral} and \ref{differentiation} below).

\subsection{Constructing the $\la$-distributions from quadrature data}\label{lambdadistr}
In a recent paper \cite{Lahti} it was shown, that the $Q$-function of a state $\rho$ can be constructed from the rotated quadrature distributions $W^{\textsf{qd}}_\rho$, by means of a generalized Markov kernel. That is, for each $z=\frac{1}{\sqrt{2}}(q+ip)$, $(q,p)\in\R^2$, there exists a function $M^{q,p}_{0}: \R\times [0,2\pi)\rightarrow\R $, such that 
$$
\langle z\vert \rho \vert z\rangle =\int_0^{2\pi} \int_\R M^{q,p}_{0} (x,\theta) W^{\textsf{qd}}_\rho (x,\theta) \, dxd\theta,
$$
for all states $\rho$. In \cite{Pellonpaa}, the generalized Markov  kernel for the $\la$-distributions was constructed. We will briefly recall the results.

First, define $W^\la_\rho (r,\theta )=:W^\la_\rho (z) =:W^\la_\rho (q,p)$, where $z=re^{i\theta}=\frac{1}{\sqrt{2}} (q+ip)$. We are interested in finding a kernel $M^{q,p}_\la$, such that we can obtain the $\la$-distributions from the quadrature distributions by integrating:
\begin{equation}\label{Markov}
W^\la_\rho (q,p) =\int_0^{2\pi} \int_\R M^{q,p}_\la (x,\theta) W^{\textsf{qd}}_\rho(x,\theta)\frac{dxd\theta}{2\pi},
\end{equation}
where $W^{\textsf{qd}}_\rho$ is the probability density related to the quadrature observable as before. It is sufficient to show the validity of the equation for coherent states $\rho=\vert \alpha\rangle\langle \alpha\vert$, $\alpha\in\C$, in which case
\begin{eqnarray}\label{koherenttijakauma}
W^\la_{\vert \alpha\rangle\langle \alpha\vert } (z) &=& \langle\alpha \vert D(z) K^\la D(z)^\ast \vert \alpha\rangle =  \langle \alpha -z\vert K^\la \vert\alpha-z\rangle =(1-\la) \sum_{k=0}^\infty \la^k\vert \langle\alpha- z\vert k\rangle\vert^2 \nonumber\\
&=& (1-\la) e^{-\vert \alpha -z\vert^2} \sum_{k=0}^\infty \la^k \frac{\vert \alpha -z\vert^{2k}}{k!} = (1-\la) e^{-(1-\la)\vert \alpha -z\vert^2}.
\end{eqnarray}

Now assume first that $\la\in\R$, $\vert\la\vert<1$ and define the function $M^{0,0}_\la :\R\times[0,2\pi) \rightarrow \C$ by 
$$
M^{0,0}_\la (x,\theta):= \frac{1+\la}{1-\la} \, Y\left(\sqrt{\frac{1+\la}{1-\la}} x \right),
$$
where $Y (x) =2\frac{d}{dx}\,{\rm daw}(x)$ as before. Define the functions $M^{q,p}_\la :\R\times [0,2\pi)\rightarrow \C$ by
$$
M^{q,p}_\la (x,\theta):=M^{0,0}_\la (x- q\cos\theta -p\sin\theta, \theta).
$$
The function $M^{q,p}_\la$ can be represented as a series of Hermite polynomials \cite{Pellonpaa}
\begin{equation}\label{kernel}
M^{q,p}_\la (x,\theta) = (1-\la )\sum_{k=0}^\infty \frac{(\la -1)^k k!}{ 2^k (2k)!} H_{2k}(x-a),
\end{equation}
where $a:=q\cos\theta +p\sin\theta=\sqrt{2}\,\textrm{Re} (ze^{-i\theta})$, $z=\frac{1}{\sqrt{2}}(q+ip)$, and the series converges absolutely. Since the absolute convergence of the series in equation \eqref{kernel} does not depend on $\la$ being real, the function $M^{q,p}_\la$ can be defined for all $\la\in\C$, $\vert \la\vert <1$.

Defining $\tilde{u}:=\sqrt{2}\, \textrm{Re} (\alpha e^{-i\theta})$, we have
$$
\int_X W^{\textsf{qd}}_{\vert \alpha\rangle\langle\alpha\vert} (x,\theta)dx =\langle\alpha \vert \mathsf{Q}_{\theta} (X)\vert\alpha\rangle =\langle \alpha e^{-i\theta} \vert \mathsf{Q}  (X)\vert\alpha e^{-i\theta}\rangle =\frac{1}{\sqrt{\pi}} \int_X e^{-(x-\tilde{u})^2} dx,
$$
for all $X\in\hB (\R)$, which implies that
$$
\int_0^{2\pi} \int_\R M^{q,p}_\la (x,\theta ) W^{\textsf{qd}}_{\vert\alpha\rangle\langle\alpha\vert}(x,\theta )\frac{dxd\theta}{2\pi} =\frac{1}{\sqrt{\pi}}\int_0^{2\pi} \int_\R M^{q,p}_\la (x,\theta) e^{-(x-\tilde{u})^2}\frac{dxd\theta}{2\pi}.
$$

Following the calculations in \cite{Lahti}, we get
\begin{eqnarray*}
&&\frac{1}{\sqrt{\pi}} \int_0^{2\pi} \int_\R M^{q,p}_\la (x,\theta) e^{-(x-\tilde{u})^2} \frac{dxd\theta }{2\pi} \\
&=& \frac{(1-\la)}{\sqrt{\pi}}\sum_{k=0}^\infty \int_0^{2\pi} \frac{(\la- 1)^k k!}{2^k  (2k)!} \int_\R H_{2k} (x)e^{-(x+a-\tilde{u})^2}\, dx\frac{d\theta}{2\pi}\\
&=&(1-\la)\sum_{k=0}^\infty \frac{(\la-1)^k k!}{(2k)!} \sum_{l=0}^{2k} { 2k \choose l} (\alpha -z)^l \overline{(\alpha-z)}^{2k-l}\int_0^{2\pi}e^{2i\theta(k-l)}\frac{d\theta}{2\pi}\\
 &=& (1-\la)\sum_{k=0}^\infty \frac{(\la-1)^k \vert \alpha -z\vert^{2k}}{k!} \\
&=&(1-\la)\sum_{k=0}^\infty e^{-(1-\la) \vert \alpha -z\vert^2},
\end{eqnarray*}
which shows that
$$
W^\la_{\vert \alpha\rangle\langle \alpha\vert } (z) = \frac{1}{\sqrt{\pi}} \int_0^{2\pi} \int_\R M^{q,p}_\la (\theta,x) e^{-(x-\tilde{u})^2} \frac{d\theta dx}{2\pi}.
$$
Thus, equation \eqref{Markov} holds for all states.

\subsection{Shifting the $\la$-parameter}
Suppose that for a given $\la\in(-1,1)$ one has obtained the $\la$-distribution $W^\la_\rho$ of some state $\rho$. One might be interested in finding the distributions for a different value, say $\la'$. It turns out that in some cases this shifting of the $\la$-parameter is needed for reconstructing the state. Indeed, the reconstruction formulas we derive in the next sections are only valid for certain values of $\la$. Therefore, we present here the formulas for this shifting.

First note that $W^\la_\rho\in L^1(\R^2)$ for all states $\rho$ and $\la\in\C$, $\vert\la\vert <1$ (see e.g. \cite[Lemma 3.1]{Werner}). This implies that the Fourier transform of $W^\la_\rho$, as well as its convolutions with other integrable functions, are well defined. Now let $\la\in(-1,1)$. According to equation \eqref{koherenttijakauma} we have for a coherent state $\vert \alpha\rangle\langle\alpha\vert$, $\alpha\in\C$
$$
W^\la_{\vert \alpha\rangle\langle \alpha\vert } (z) = (1-\la) e^{-(1-\la)\vert \alpha -z\vert^2}.
$$
Putting $\alpha=\frac{1}{\sqrt{2}} (x+iy)$ and $z=\frac{1}{\sqrt{2}}(q+ip)$, we get 
$$
W^\la_{\vert \alpha\rangle\langle \alpha\vert } (q,p) =(1-\la) e^{-\frac{(1-\la)}{2}[(x-q)^2 +(y-p)^2]}.
$$
Let $\la'\in(-1,1)$, $\la'>\la$, and define the function $g_{\la,\la'}:\R^2\rightarrow \C$ by
$$
g_{\la,\la'}(q,p)=\frac{1}{2\pi} \frac{(1-\la')(1-\la)}{\la'-\la}e^{-\frac{1}{2} \frac{(1-\la')(1-\la)}{\la'-\la} (q^2+p^2)},
$$
so that $g_{\la,\la'}\in L^1(\R^2)$ for all $\la,\la'$. Now a direct calculation shows that the distribution $W^{\la'}_{\vert\alpha\rangle\langle\alpha\vert}$ is the convolution of $W^\la_{\vert\alpha\rangle\langle\alpha\vert}$ with the function $g_{\la,\la'}$, that is 
\begin{eqnarray*}
&&(W^\la_{\vert\alpha\rangle\langle\alpha\vert}\ast g_{\la,\la'})(u,v) = \int_{\R^2} W^\la_{\vert\alpha\rangle\langle\alpha\vert} (u-q,v-p) g_{\la,\la'} (q,p)\,dqdp\\
&=&\frac{1}{2\pi} \frac{(1-\la')(1-\la)^2}{\la'-\la} \int_{\R^2} e^{-\frac{(1-\la)}{2} [(q+x-u)^2 +(p+y-v)^2]} e^{-\frac{1}{2}\frac{(1-\la')(1-\la)}{\la'-\la} (q^2 +p^2)} \, dqdp\\
&=& (1-\la') e^{-\frac{(1-\la')}{2}[ (x-u)^2 +(y-v)^2]} = W^{\la'}_{\vert\alpha\rangle\langle\alpha\vert} (u,v),
\end{eqnarray*}
for all $(u,v)\in\R^2$. Since this is valid for all coherent states, we have $W^{\la'}_\rho = W^\la_\rho\ast g_{\la,\la'}$, for all states $\rho$ and $\la,\la'\in(-1,1)$, $\la' >\la$.

The $\la$-distributions obtainable via measurements are those, for which $\la\geq 0$, so it is of greater interest to find the inverse for the relation above. For this, we need to be able to invert the convolution transform, which of course puts some restrictions on the distributions in question. Let $\la$ and $\la'$ be as before. Then, by the Fourier theory, we have $\hat{W}^{\la'}_\rho = \widehat{W^\la_\rho\ast g_{\la,\la'}} = 2\pi \, \hat{W}^\la_\rho \cdot \hat{g}_{\la,\la'}$. The Fourier transform of $g_{\la,\la'}$ can easily be computed, and we get
$$
\hat{g}_{\la,\la'}(u,v) =\frac{1}{2\pi} e^{-\frac{1}{2} \frac{\la'-\la}{(1-\la')(1-\la)} (u^2 +v^2)},
$$
for all $(u,v)\in\R^2$, showing that $\hat{g}_{\la,\la'}$ is pointwise nonzero. Thus, we have $\hat{W}^\la_\rho =\frac{1}{2\pi} \frac{\hat{W}^{\la'}_\rho}{\hat{g}_{\la,\la'}}$. If $\frac{1}{2\pi} \frac{\hat{W}^{\la'}_\rho}{\hat{g}_{\la,\la'}}\in L^1(\R^2)$, we have
\begin{eqnarray}\label{konvoluutiokaava}
W^\la_\rho (q,p) &=&\frac{1}{(2\pi)^2} \int_{\R^2} e^{i(qu+pv)} \frac{\hat{W}^{\la'}_\rho(u,v)}{\hat{g}_{\la,\la'}(u,v)}\, dudv \nonumber\\
&=&\frac{1}{2\pi} \int_{\R^2} e^{i(qu+pv)} e^{\frac{1}{2} \frac{\la'-\la}{(1-\la')(1-\la)}(u^2+v^2)}\hat{W}^{\la'}_\rho(u,v)\, dudv 
\end{eqnarray}
for almost all $(q,p)\in\R^2$. Note that the above integrability condition can, at least in principle, be tested directly on the measured distribution, and thus no \textit{a priori} information on the state is needed. It is easy to see that there actually exist states for which the condition is satisfied. The significance of equation \eqref{konvoluutiokaava} is that if one measures $W^{\la'}_\rho$ for some $\la'\geq 0$, then the distributions for other values of $\la$, may be calculated, provided that the integral exists. In particular, the $Q$-function and the distributions corresponding to the negative values of $\la$ can be obtained in this way.

\subsection{Reconstruction via integration}\label{integral}
We will first derive a reconstruction formula which can be obtained by integrating the $\la$-distribution with respect to suitable averaging functions. The formula can be seen as a direct consequence of the overlap relation for the $\la$-distributions \cite{Cahill1969} (the overlap relation has also been treated in e.g. \cite[pp. 58-59]{Leonhardt}). A different form of the formula appears also in \cite{Mancini1997}. However, we will go through the mathematical details to ensure the validity of the formula. As it turns out, the formula only works for negative values of the parameter $\la$.

Let $\la\in (-1,0)$, and for each $n,m\in\N$, let $K^\la_{nm}:[0,\infty)\rightarrow \C$ be the function defined by equation \eqref{eq5}. Substituting $\la$ with $\la^{-1}$ in equation \eqref{eq5} we see that also the functions $K^{\la^{-1}}_{nm}$ are well defined for each $n,m\in\N$. We wish to integrate the function $W^\la_{\rho,k}(r) K^{\la^{-1}}_{l,l+k}(r)$ with respect to $r$, and in order to do that we need to pay attention to some convergence issues. This is done in the following lemma (for the proof, see Appendix A). 
\begin{lemma}\label{integraalilemma}
 Let $\rho$ be a state and $K_{n,n+l}^\la$ be the function defined in equation \eqref{eq5}. Then
$$
\int_0^\infty W^\la_{\rho,k}(r) K^{\la^{-1}}_{l,l+k}(r)r\,dr = \sum_{n=0}^\infty \rho_{n+k,n} \int_0^\infty K^{\la}_{n,n+k}(r)K^{\la^{-1}}_{l,l+k}(r)r\,dr
$$
for all $k\in\N$, $\la\in(-1,0)$.  
\end{lemma}

Using the orthogonality relation 
$$
\int_0^\infty e^{-x} x^\alpha L^\alpha_n (x) L^\alpha_m (x)\,dx =\frac{\Gamma[\alpha +n +1]}{n!}\delta_{nm}
$$
for the associated Laguerre polynomials \cite[p. 844, 7.414(3)]{Gradshteyn}, we calculate
\begin{eqnarray*}
 &&\int_0^\infty W^\la_{\rho,k} (r)K^{\la^{-1}}_{l,l+k}(r) r\,dr  = \sum_{n=0}^\infty \rho_{n+k,n} \sqrt{\frac{n!l!}{(n+k)!(l+k)!}}(1-\la)^{k+1}(1-\la^{-1})^{k+1} \la^n \la^{-l}  \\
&&\times\int_0^\infty r^{2k} e^{-(1-\la)r^2}e^{-(1-\la^{-1})r^2} L^k_n ((2-\la-\la^{-1})r^2 ) L^k_l ((2-\la-\la^{-1})r^2) r\,dr\\
&=&\frac{1}{2}\sum_{n=0}^\infty \rho_{n+k,n} \sqrt{\frac{n!l!}{(n+k)!(l+k)!}} \la^{n-l}\int_0^\infty  x^k e^{-x} L^k_n(x) L^k_l (x)\,dx \\
&=&\frac{1}{2} \sum_{n=0}^\infty \rho_{n+k,n} \sqrt{\frac{n!l!}{(n+k)!(l+k)!}}  \la^{n-l}\frac{\Gamma [k+n+1]}{n!} \delta_{ln} \\
&=&\frac{1}{2}\rho_{l+k,l} \frac{l!}{(l+k)!}  \frac{(l+k)!}{l!} \\
 &=&\frac{1}{2} \rho_{l+k,l}
\end{eqnarray*}
for each $k,l\in\N$. This corresponds to equation \eqref{integration}, where the series on the right-hand side has been reduced to trivial. Hence we can immediately write the reconstruction formula for the matrix elements:
\begin{equation}\label{integraalikaava}
\rho_{n+k,n} =2 \int_0^\infty W^\la_{\rho,k}(r) K^{\la^{-1}}_{n,n+k}(r) r\,dr.
\end{equation}
Note that this method works only for negative values of $\la$, and therefore the state can not be directly reconstructed from measurement statistics. However, as we have shown in the previous sections, the required distributions can be constructed from measurement data, at least in some cases.

To further illustrate the fact that the negativity of the $\la$-parameter is in fact a necessary condition for the formula, we give the following counter example. Consider the vacuum state $\rho =\vert 0\rangle\langle 0\vert$, for which we have
$$
W^\la_{\vert 0\rangle\langle 0\vert}(r,\theta) =(1-\la )e^{-(1-\la)r^2}.
$$
If we attempt to reconstruct the only nonzero matrix element $\rho_{00}$ by directly using equation \eqref{integraalikaava}, we find that
$$
\rho_{00} = 2(1-\la)(1-\la^{-1}) \int_0^\infty e^{-(1-\la)(1-\la^{-1})r^2}r\, dr, 
$$
which clearly diverges for $\la\in(0,1)$, since in that case $(1-\la)(1-\la^{-1})\leq 0$. Thus, the formula works for {\em all} states only if $\la\in(-1,0)$. In order to obtain a valid formula for other values of $\la$, we need to use a different method.

\subsection{Reconstruction via differentiation}\label{differentiation}
We now proceed to the reconstruction formula obtained via differentiation. We will start with a lemma, which is needed for inverting the infinite matrix identity \eqref{eq9} appearing in the derivation of the reconstruction formula. The proof can again be found in Appendix A.

\begin{lemma}\label{kertomamatriisilemma}
Let $(x_n)_{n\in\N}$ be a sequence of complex numbers such that
$$
\sum_{n=p}^\infty \frac{x_{n}}{(n-p)!}=y_p\in\C,\hspace{1cm}p\in\N.
$$
\begin{itemize}
\item[(a)] For each $n\in \N$, denote
$$
R^n_k:=\sum_{n'=k+1}^\infty \sum_{p=n}^k (-1)^{p+n}\frac{x_{n'}}{(p-n)!(n'-p)!}, \ \ k\geq n.
$$
Then
$$
\sum_{p=n}^\infty \frac{(-1)^{p-n}y_{p}}{(p-n)!}=x_n,
$$
if and only if $\lim_{k\rightarrow\infty} R^n_k=0$.
\item[(b)] Suppose that $\lim_{n\rightarrow\infty} |x_n|=0$. Then $\lim_{k\rightarrow\infty} R_k^n=0$
for all $n\in \N$.
\end{itemize}
\end{lemma}

With the previous notations we have
\begin{equation}\label{sarja1}
e^{(1-\la)r^2}W^\la_{\rho,k}(r)=\sum_{n=0}^\infty \rho_{n+k,n}e^{(1-\la)r^2}K^\la_{n,n+k}(r),
\end{equation}
where the series converges absolutely and uniformly on any finite interval $[0,R)$. In order to be able
to differentiate the series term by term, and evaluate the value at the origin, we must prove
that the series for each derivative converges uniformly in an interval $[0,\epsilon)$, where $\epsilon>0$. This is done in the following lemma (for the proof, see Appendix A).

\begin{lemma}\label{derivaattalemma}
Let $\rho$ be a state and $\la\in\C$, $\vert\la\vert<1$. Then there exists an $\epsilon >0$ such that the series
\begin{equation}\label{derivaattasarja}
\sum_{n=0}^\infty \rho_{n+k,n}\frac{\pa^l [e^{(1-\la)r^2}K^\la_{n,n+k}(r)]}{\pa r^l}
\end{equation}
converges uniformly on $[0,\epsilon)$ for each $l\in\N$.
\end{lemma}

The series in equation \eqref{sarja1} can thus be differentiated termwise. We then get
$$
W^\la_{\rho,k,l}:=\frac{\pa^l [e^{(1-\la)r^2}W^\la_{\rho,k}(r)]}{\pa r^l}\bigg|_{r=0}
=\sum_{n=0}^\infty \rho_{n+k,n}\frac{\pa^l [e^{(1-\la)r^2}K^\la_{n,n+k}(r)]}{\pa r^l}\bigg|_{r=0},
$$
which corresponds to equation \eqref{differentiation}. Now 
$$
\frac{\pa^l r^{2u+k}}{\pa r^l}\bigg|_{r=0}=l!\delta_{2u+k,l}
$$
so that we get two (nonzero) cases: (i) when $k$ is even, then $l$ must be even, and 
(ii) when $k$ is odd, then $l$ must be odd.

Let $h,\,j\in\N$ and consider case (i) when $k=2h$ and $l=2j$.
Now
\begin{eqnarray}\nonumber
W^\la_{\rho,2h,2j}&=&\sum_{n=0}^\infty \rho_{n+2h,n}
\sqrt{\frac{n!}{(n+2h)!}}(1-\la)^{2h+1}
\sum_{u=0}^n\frac{\la^{n-u}}{u!}{{n+2h}\choose{n-u}}(1-\la)^{2u}(2j)!\delta_{u,j-h} \\
&=&\nonumber
\sum_{n=\max\{0,j-h\}}^\infty \rho_{n+2h,n}
\sqrt{\frac{n!}{(n+2h)!}}
\frac{(1-\la)^{2j+1}\la^{n-j+h}(n+2h)!(2j)!}{(j-h)!(n-j+h)!(j+h)!}\\
&=&
{2j \choose j-h}{(1-\la)^{2j+1}}\la^{h-j}
\sum_{n=\max\{0,j-h\}}^\infty\frac{\rho_{n+2h,n}
\la^{n}\sqrt{{(n+2h)!}n!}}{(n-j+h)!}.\label{eq7}
\end{eqnarray}
Consider then case (ii) and put  $k=2h+1$ and $l=2j+1$. Then
\begin{equation}\label{eq8}
W^\la_{\rho,2h+1,2j+1}={2j+1 \choose j-h}{(1-\la)^{2j+2}}\la^{h-j}\sum_{n=\max\{0,j-h\}}^\infty 
\frac{\rho_{n+2h+1,n}\la^{n}
\sqrt{(n+2h+1)!n!}}{(n-j+h)!}.
\end{equation}
Define, for all $n,\,m\in\N$,
$$
\tilde \rho^\la_{mn}:=\rho_{mn}\la^{\min\{m,n\}}\sqrt{m!n!}.
$$
Then both equations \eqref{eq7} and \eqref{eq8} reduce to the single one:
\begin{equation}\label{eq9}
\sum_{n=\max\{0,(l-k)/2\}}^\infty\frac{\tilde \rho^\la_{n+k,n}}{\big(n-(l-k)/2\big)!}=
W^\la_{\rho,k,l}{l \choose (l-k)/2}^{-1}{(1-\la)^{-(l+1)}}\la^{(l-k)/2}
\end{equation}
where either $k=2h$ and $l=2j$ or $k=2h+1$ and $l=2j+1$.

In order to invert equation \eqref{eq9}, we need to check that the necessary and sufficient condition of Lemma \ref{kertomamatriisilemma} is satisfied. It turns out that there are two different cases. If $\vert\la\vert <\frac{1}{2}$, then the equation may be inverted regardless of the state in question. If $\la\geq\frac{1}{2}$, then one may need some prior knowledge of the state, namely, that the condition of Lemma \ref{kertomamatriisilemma} is satisfied.

Using the identity $\sum_{k=0}^m (-1)^k {n \choose k} = (-1)^m {n-1 \choose m}$, which holds for $n\geq 1$ \cite[p. 3, 0.151(4)]{Gradshteyn}, we get by straightforward calculation
\begin{eqnarray*}
R^n_k &=&\sum_{n'=k+1}^\infty \sum_{p=n}^k (-1)^{p+n} \frac{\sqrt{(n'+l)!n'!}}{(p-n)!(n'-p)!} \la^{n'} \rho_{n'+l,n'}\\
&=&\sum_{n'=k+1}^\infty (-1)^{k+n} \la^{n'} \rho_{n'+l,n'} \frac{\sqrt{(n'+l)!n'!}}{(n'-n)(n'-k-1)!(k-n)!}\\
&=&\sum_{v=1}^\infty (-1)^{k+n} \la^{v+k} \rho_{v+l+k,v+k} \frac{\sqrt{(v+l+k)!(v+k)!}}{(v+k-n)(v-1)!(k-n)!}.
\end{eqnarray*}
Since $\vert \rho_{v+l+k,v+k}\vert \leq 1$ and $n\leq k$, we get the bound
\begin{eqnarray*}
\vert R^n_k \vert &\leq& \frac{\vert \la\vert^k}{(k-n)!}\sum_{v=1}^\infty  \vert\la\vert^{v} \frac{\sqrt{(v+l+k)!(v+k)!}}{v!} \\
&\leq& \frac{(k+l)!\vert \la\vert^k}{(k-n)!}\sum_{v=1}^\infty  \vert\la\vert^{v} { v+l+k \choose v}\\
&\leq& \frac{(k+l)!\vert \la\vert^k}{(k-n)!}\frac{1}{(1-\vert\la\vert)^{l+k+1}}\\
&=&\frac{(k+l)!}{(k-n)!}\frac{1}{(1-\vert\la\vert)^{l+1}}\left(\frac{\vert\la\vert}{1-\vert\la\vert}\right)^k,
\end{eqnarray*}
where we have used the identity $\frac{1}{(1-z)^{\alpha +1}} =\sum_{n=0}^\infty { n+\alpha \choose n} z^n$, valid when $\vert z\vert < 1$. If $\vert\la\vert <\frac{1}{2}$, then $\frac{\vert\la\vert}{1-\vert \la\vert} <1$, and thus $\lim_{k\rightarrow \infty} R^n_k =0$ for all $n\in\N$.

Now Lemma \ref{kertomamatriisilemma} may be used on equation \eqref{eq9} to get
$$
\tilde \rho^\la_{n+k,n}=\sum_{p=n}^\infty\frac{(-1)^{p-n}}{(p-n)!}
W^\la_{\rho,k,2p+k}{2p+k \choose p}^{-1}{(1-\la)^{-(2p+k+1)}}\la^{p}.
$$
Thus, we get the reconstruction formula
\begin{eqnarray}\label{derivointikaava}
\rho_{n+k,n}&=&\frac{1}{\sqrt{(n+k)!n!}}
\sum_{p=n}^\infty{2p+k \choose p}^{-1}\frac{(-1)^{p-n}\la^{p-n}}{(1-\la)^{2p+k+1}(p-n)!}
W^\la_{\rho,k,2p+k} \nonumber\\
&=&
\sqrt{\frac{n!}{(n+k)!}}\sum_{p=n}^\infty{p\choose n}\frac{(p+k)!}{(2p+k)!}
\frac{(-\la)^{p-n}}{(1-\la)^{2p+k+1}}
\frac{\pa^{2p+k} [e^{(1-\la)r^2}W^\la_{\rho,k}(r)]}{\pa r^{2p+k}}\bigg|_{r=0} 
\end{eqnarray}
which holds for all $n,\,k\in\N$, at least if $\vert\la\vert <\frac{1}{2}$. A similar bound for the parameter has also appeared in previous works concerning state reconstruction from the $\la$-distributions, see e.g. \cite{Leonhardt1993, Macchiavello1995}.

In this case the reconstruction formula \eqref{derivointikaava} works also for the $Q$-function. In fact, defining $Q_\rho(z):=\langle z\vert \rho\vert z\rangle =W^0_\rho(r,\theta)$, where $z=re^{i\theta}$, we get
\begin{equation}
\rho_{n+k,n} = \frac{\sqrt{n!(n+k)!}}{(2n+k)!}\frac{\pa^{2n+k}}{\pa r^{2n+k}} \left[e^{r^2}\frac{1}{2\pi}\int_0^{2\pi}e^{ik\theta} Q_\rho(re^{i\theta}) d\theta\right]\bigg|_{r=0}. 
\end{equation}
\

\begin{remark}\rm We close this section with a note on reconstructing the state from the Wigner function. In the above discussion we have not made any assumptions on the state of the system, and as a result, the values of the parameter $\la$ had to be restricted. In particular, neither of the methods gives us a reconstruction formula for the Wigner function, that is, for $\la=-1$. Since in both cases the restrictions are related to the convergence issues, it is clear that one can obtain a reconstruction formula for the Wigner function if one knows \textit{a priori} that the matrix elements of the state fall off fast enough. For example, if one assumes that the series $\sum_{n=0}^\infty \vert \rho_{n+k,n}\vert \sqrt{\frac{(n+k)!}{n!}}$ converges for all $k\in\N$, the integration method gives a formula also for the Wigner function. Clearly the coherent state $\rho=\vert z\rangle\langle z\vert$ satisfies this condition. It is easy to check that the differentiation method also works for the Wigner function of a coherent state. That is, the series \eqref{derivaattasarja} converges for $\la=-1$  and the condition of Lemma \ref{kertomamatriisilemma} is satisfied. Another assumption which allows the reconstruction of the state for all (nonzero) values of $\la$, is the finiteness of the state matrix. In that case, the differentiation method works well. It is interesting to note that the method works regardless of the size of the state matrix, as long as it is finite. A similar observation can be made about the integration method in the case of the Wigner function.
\end{remark}
\

\section{Conclusion}
We have considered the problem of state reconstruction in the cases where the measured distributions are either rotated quadrature distributions or $\la$-parametrized distributions. We have proved three reconstruction formulas using the method of infinite matrix inversion, while paying close attention to the mathematical details. In particular, we have analyzed the convergence issues which, in the case of the $\la$-distributions, give restrictions concerning the values of $\la$ for which the reconstruction formulas are valid. For the quadrature distributions we have given an alternative proof for the known formula \eqref{reconstruction}. In the case of the $\la$-distributions we derived two different formulas, namely equations \eqref{integraalikaava} and \eqref{derivointikaava}, obtained via two different methods. The integration method provides a formula, which works only for negative values of $\la$, and therefore cannot be used directly on measurement statistics. However, as we have seen, there are ways to go around this problem. The $\la$-distributions for the negative values of $\la$ can be constructed either from the rotated quadrature distributions by means of a generalized Markov kernel, or from another $\la$-distribution, corresponding to a positive $\la$, by inverting a certain convolution transform. The differentiation method has also its limitations with $1/2$ as the upper bound for $\vert\la\vert$.

\

\noindent {\bf Acknowledgment.} We wish to thank Pekka Lahti for useful comments on the manuscript. J. K. was supported by Emil Aaltonen Foundation and Finnish Cultural Foundation during the preparation of the manuscript. J. S. was supported by Turku University Foundation.

\section*{Appendix A}
We have collected here the proofs of the three technical lemmas needed in section \ref{lambda}, namely Lemmas \ref{integraalilemma}, \ref{kertomamatriisilemma}, and \ref{derivaattalemma}.

\

\noindent\textbf{Proof of Lemma \ref{integraalilemma}.} Let $k\in\N$ and define $x:=(2-\la-\la^{-1})r^2$. Define for each $N\in\N$ the function $f^\la_{N,k,l}:[0,\infty)\rightarrow\C$ by
$$
f^\la_{N,k,l} (x) :=\frac{1}{2} \sqrt{\frac{l!}{(l+k)!}}  \la^{-l} \sum_{n=0}^N  \rho_{n+k,n} \sqrt{\frac{n!}{(n+k)!}}  \la^n  x^k e^{-x} L^k_n (x) L^k_l (x).
$$
Now the function $ x^k e^{-x} L^k_n (x) L^k_l (x)$ can be estimated \cite{Duran}, and we get 
$$
\vert x^k e^{-x} L^k_n (x) L^k_l (x)\vert \leq 2^k (n+1)\cdots (n+k) e^{-\frac{x}{2}}\vert L^k_l(x)\vert=2^k \frac{(n+k)!}{n!}e^{-\frac{x}{2}} \vert L^k_l (x)\vert
$$
for all $x\in[0,\infty)$. This implies
\begin{eqnarray*}
\vert f^\la_{N,k,l} (x)\vert &=&\frac{1}{2} \sqrt{\frac{l!}{(l+k)!}}  \vert\la\vert^{-l} \bigg\vert\sum_{n=0}^N  \rho_{n+k,n} \sqrt{\frac{n!}{(n+k)!}}  \la^n  x^k e^{-x} L^k_n (x) L^k_l (x)\bigg\vert \\
&\leq&\frac{1}{2} \sqrt{\frac{l!}{(l+k)!}} \vert\la\vert^{-l} \sum_{n=0}^N  \vert \rho_{n+k,n}\vert \sqrt{\frac{n!}{(n+k)!}}  \vert\la\vert^n  \vert x^k e^{-x} L^k_n (x) L^k_l (x)\vert \\
&\leq&\frac{1}{2} \sqrt{\frac{l!}{(l+k)!}}  \vert\la\vert^{-l} \sum_{n=0}^N  \vert \rho_{n+k,n}\vert \sqrt{\frac{n!}{(n+k)!}}  \vert\la\vert^n 2^k \frac{(n+k)!}{n!} e^{-\frac{x}{2}} \vert L^k_l (x)\vert\\
&\leq& 2^{k-1} \sqrt{\frac{l!}{(l+k)!}}  \vert\la\vert^{-l} \sum_{n=0}^N  \vert \rho_{n+k,n}\vert  \vert\la\vert^n \sqrt{\frac{(n+k)!}{n!}} e^{-\frac{x}{2}} \vert L^k_l (x) \vert.
\end{eqnarray*}
Since $\vert \rho_{n+k,n}\vert\leq 1$ for all $n,k\in\N$, the series $\sum_{n=0}^\infty \vert \rho_{n+k,n}\vert \vert\la\vert^n\sqrt{\frac{(n+k)!}{n!}}$ converges for all $k\in\N$ by the ratio test. That is, for each $k\in\N$, there exists a $C_k \geq 0$ such that $\sum_{n=0}^N  \vert \rho_{n+k,n}\vert  \vert\la\vert^n \sqrt{\frac{(n+k)!}{n!}}\leq C_k$, for all $N\in\N$. Thus we have
$$
\vert f^\la_{N,k,l} (x)\vert \leq 2^{k-1} \sqrt{\frac{k!}{(l+k)!}}  \vert\la\vert^{-l} C_k e^{-\frac{x}{2}}\vert L^k_l (x) \vert,
$$
for all $N\in\N$. Since $L^k_l (x)$ is a polynomial of order $l$, the function $e^{-\frac{x}{2}}\vert L^k_l (x)\vert$ is integrable over $[0,\infty)$, so the claim follows from the dominated convergence theorem.
\begin{flushright}$\square$\end{flushright}

\

\noindent\textbf{Proof of Lemma \ref{kertomamatriisilemma}.} Fix $n\in \N$, and consider the partial sum
$$
S_k :=\sum_{p=n}^k \frac{(-1)^{p-n}y_{p}}{(p-n)!}.
$$
Substitute the convergent series $y_p=\sum_{n'=p}^\infty \frac{x_{n'}}{(n'-p)!}$ into it:
\begin{eqnarray*}
\sum_{p=n}^k \frac{(-1)^{p-n}}{(p-n)!}\sum_{n'=p}^\infty \frac{x_{n'}}{(n'-p)!}&=&
\sum_{p=n}^k \sum_{n'=p}^\infty(-1)^{p+n}{p\choose n}{n'\choose p}\frac{n!}{n'!}x_{n'}
=\sum_{p=0}^k \sum_{n'=0}^\infty(-1)^{p+n}{p\choose n}{n'\choose p}\frac{n!}{n'!}x_{n'}\\
&=&
\sum_{n'=0}^\infty\left[\sum_{p=0}^k
(-1)^{p+n}{n'\choose p}{p\choose n}
\right]\frac{n!}{n'!}x_{n'}=x_n +R_k^n
\end{eqnarray*}
where we have used the result $\sum_{p=0}^k
(-1)^{p+n}{n'\choose p}{p\choose n}=\delta_{n,n'}$, which holds when $k\geq n'$ \cite[p. 4]{Riordan}.
We have proved (a). To prove (b), we use a crude estimate
\begin{eqnarray*}
|R_k^n|&\leq& \sum_{n'=k+1}^\infty \sum_{p=n}^k \frac{|x_{n'}|}{(p-n)!(n'-p)!} \leq 
\sum_{n'=k+1}^\infty \sum_{p=n}^k \frac{|x_{n'}|}{(p-n)!(n'-k)!} \\
&\leq& e \sum_{n'=k+1}^\infty\frac{|x_{n'}|}{(n'-k)!}
= e \sum_{n'=1}^\infty\frac{|x_{n'+k}|}{n'!}.
\end{eqnarray*}
Since $\lim_{n\rightarrow \infty} |x_{n}| =0$, the sequence $(x_n)_{n\in\N}$ is bounded, i.e. there is a $K>0$ with $|x_n|\leq K$ for all $n\in \N$. Since $\sum_{n'=1}^\infty\frac{K}{n'!}<\infty$,  we can use e.g. the dominated convergence theorem to get
$$
0=\sum_{n'=1}^\infty 0 = \sum_{n'=1}^\infty \lim_{k\rightarrow \infty} \frac{|x_{n'+k}|}{n'!}
= \lim_{k\rightarrow \infty}\sum_{n'=1}^\infty  \frac{|x_{n'+k}|}{n'!},
$$
which proves that $\lim_{k\rightarrow\infty} R_k^n =0$. The proof is complete.
\begin{flushright}$\square$\end{flushright}

\

\noindent\textbf{Proof of Lemma \ref{derivaattalemma}.}
From eqs.\ \eqref{eq5} and \eqref{eq6} we see that
\begin{eqnarray*}
\frac{\pa^l [e^{(1-\la)r^2}K^\la_{n,n+k}(r)]}{\pa r^l}&=&
\sqrt{\frac{n!}{(n+k)!}}(1-\la)^{k+1}
\sum_{u=0}^n\frac{\la^{n-u}}{u!}{{n+k}\choose{n-u}}(1-\la)^{2u}\frac{\pa^l r^{2u+k}}{\pa r^l}.
\end{eqnarray*}
First, suppose that $\la\neq 0$. Since $\vert\la\vert <1$, we can choose positive numbers $\epsilon$ and $\delta$  such that $(1 +\delta)\vert\la\vert <1$ and $\vert (1-\la)^2\la^{-1}\vert \epsilon^2\leq \delta$. 
Let $u_0$ be the smallest natural number with $2u_0\geq l-k$. Then for $u\geq u_0$ we have
$$
\frac{\pa^l r^{2u+k}}{\pa r^l} =(2u+k )(2u+k-1)\cdots (2u+k-l+1) r^{2u+k-l} \leq (2u+k)^l \epsilon^{2u+k-l}
$$
for all $r\in [0,\epsilon)$. Thus we have
\begin{eqnarray*}
\bigg|\frac{\pa^l [e^{(1-\la)r^2}K^\la_{n,n+k}(r)]}{\pa r^l}\bigg| &\leq &\sqrt{\frac{n!}{(n+k)!}} |1-\la|^{k+1}|\la|^{n}(n+k)!\sum_{u=u_0}^n\frac{|(1-\la)^2\la^{-1}|^u}{u!(n-u)!}\frac{(2u+k)^l}{(k+u)!} \epsilon^{2u+k-l}\\
&\leq&  |1-\la|^{k+1}|\la|^{n}(n+k)! \epsilon^{k-l} \sum_{u=0}^n\frac{\delta^u}{u!(n-u)!}\frac{(2u+k)^l}{(k+u)!} \\
&\leq& |1-\la|^{k+1}|\la|^{n} \frac{(n+k)}{n!}! \epsilon^{k-l} \sum_{u=0}^n { n\choose u} \delta^u\frac{(2u+k)^l}{(k+u)!}.
\end{eqnarray*}
Now the function $u\mapsto \frac{(2u+k)^l}{(k+u)!}$ is bounded by, say, $M>0$, so
$$
\bigg|\frac{\pa^l [e^{(1-\la)r^2}K^\la_{n,n+k}(r)]}{\pa r^l}\bigg| \leq  M |1-\la|^{k+1} (1+\delta)^n|\la|^n \frac{(n+k)!}{n!} \epsilon^{k-l} 
$$
for all $r\in [0,\epsilon)$. Since $(1+\delta)|\la|<1$ the series $\sum_{n=0}^\infty \frac{(n+k)!}{n!} (1+\delta)^n\vert\la\vert^n$ converges by the ratio test. The claim follows from the fact that $\vert \rho_{n+k,n}\vert\leq 1$ for all $n,k\in\N$. 

Suppose now, that $\la=0$. Let $n_0$ be the smallest natural number with $2n_0 \geq l-k$. The series \eqref{derivaattasarja} reduces to
$$
\sum_{n=n_0}^\infty \rho_{n+k,n} \frac{1}{\sqrt{n!(n+k)!}} \frac{\pa^l r^{2n+k}}{\pa r^l} =\sum_{n=n_0}^\infty \rho_{n+k,n} \frac{1}{\sqrt{n!(n+k)!}}\frac{(2n+k)!}{(2n+k-l+1)!} r^{2n+k-l},
$$
which clearly converges on any interval $[0,R)$, $R>0$. The proof is complete. 
\begin{flushright}$\square$\end{flushright}

\section*{Appendix B}
In this appendix, we prove two formulas which we used in the derivations of the reconstruction formulas. The first one gives the coefficients $c_{ln}(k)$ needed in the quadrature reconstruction formula, and the second one is the formula of Cahill and Glauber, first proved in \cite{Cahill1969}. Since the original proof does not contain all the mathematical details, we reproduce it here somewhat more carefully.

The function $Y:\R\rightarrow\R$ defined in section \ref{kvadratuuri} is an analytic function, and the derivatives of $Y$ have the series representations
\begin{eqnarray*}
Y^{(2p)}(x) &=& (-1)^p 2^p\sum_{k=0}^\infty \frac{(-1)^k (k+p)!}{2^k (2k)!}H_{2k}(x), \\
Y^{(2p+1)}(x) &=& (-1)^{p+1} 2^p\sum_{k=0}^\infty \frac{(-1)^k (k+p+1)!}{2^k (2k+1)!}H_{2k+1}(x)
\end{eqnarray*}
for all $x\in\R$ (see proof of Lemma 1 in the appendix of \cite{KLP2008}). We use the following well-known result \cite[p. 838, 7.375(1)]{Gradshteyn}: for all $m,\,n,\,l\in\N$
$$
I_{mnl}=\int_\R H_m(x)H_n(x)H_l(x)e^{-x^2}dx=
\begin{cases}
\frac{\sqrt{\pi \, 2^{m+n+l}}\;m!\,n!\,l!}{
\big(\frac12(m+n-l)\big)!\,\big(\frac12(n+l-m)\big)!\,\big(\frac12(l+m-n)\big)!}, & m+n+l \text{ is even},\\
0, & m+n+l \text{ is odd}\\
\end{cases}
$$
where the last case follows from the fact that $H_mH_nH_l$ is an odd function when $m+n+l$ is odd.
Note that $I_{mnl}=0$ (in the even case) also when $l<|n-m|$, since then either $\frac12(n+l-m)$ or $\frac12(l+m-n)$ is a negative integer.

For any $j\in\N$ the operator $Y^{(j)}(Q)$ is bounded and we can calculate its matrix elements. Let $p\in\N$ be fixed.
For all $m,\,n\in\N$
\begin{eqnarray*}
\langle m|Y^{(2p)}(Q)|n\rangle 
&=&
\frac{(-2)^p}{\sqrt{\pi\,2^{m+n}m!\,n!}}\sum_{k=0}^{[(m+n)/2]}\frac{(-1)^k(k+p)!\,I_{m,n,2k}}{2^k(2k)!},\\
\langle m|Y^{(2p+1)}(Q)|n\rangle 
&=&
\frac{-(-2)^p}{\sqrt{\pi\,2^{m+n}m!\,n!}}\sum_{k=0}^{[(m+n-1)/2]}\frac{(-1)^k(k+p+1)!\,I_{m,n,2k+1}}{2^k(2k+1)!},
\end{eqnarray*}
where $[c]$ means the integer part of $c$; e.g.\ $[1.5]=1$, and $[-0.5]=0$.
Note that $\langle m|Y^{(2p)}(Q)|n\rangle=0$ when $m+n$ is odd, and $\langle m|Y^{(2p+1)}(Q)|n\rangle=0$ when $m+n$ is even. As we will see, one gets even more zero elements.

First, let $m+n$ be even.
Since $\langle m|Y^{(r)}(Q)|n\rangle=\langle n|Y^{(r)}(Q)|m\rangle$ we may assume that
$m\le n$ and define $u'=n-m\ge 0$. Then $n=m+u'$ and
$m+n=2m+u'$ so that $u'=2u$ where $u\in\N$.
Hence, let $m,\,u\in\N$ and calculate using \cite[p. 8]{Riordan}
\begin{eqnarray*}
 \langle m|Y^{(2p)}(Q)|m+2u\rangle 
&=&
(-2)^p\sqrt{m!\,(m+2u)!}\sum_{k=u}^{m+u}\frac{(-1)^k(k+p)!}
{(m-k+u)!\,(k+u)!\,(k-u)!} \\
&=&
(-1)^{p+u}2^p\sqrt{m!\,(m+2u)!}\sum_{l=0}^{m}\frac{(-1)^l(l+u+p)!}
{(m-l)!\,(l+2u)!\,l!} \\
&=&
(-1)^{p+u}2^p(p+u)!\sqrt{\frac{m!}{(m+2u)!}}{{m+u-p-1}\choose{m}}\\
&=&
(-1)^{p+u+m}2^p(p+u)!\sqrt{\frac{m!}{(m+2u)!}}
{{p-u}\choose{m}}\\
&=&0,\;\;\;\text{ if and only if $p\ge u$ and $m>p-u$.}
\end{eqnarray*}

Secondly, let $m+n$ be odd, $m\le n$, and define $u'=n-m\ge 0$.
Then $n=m+u'$ and
$m+n=2m+u'$ so that $u'=2u+1$ where $u\in\N$.
Hence, let $m,\,u\in\N$ and calculate
\begin{eqnarray*}
&&\langle m|Y^{(2p+1)}(Q)|m+2u+1\rangle \\
&=&
-(-2)^p\sqrt{2}\sqrt{m!\,(m+2u+1)!}\sum_{k=u}^{m+u}\frac{(-1)^k(k+p+1)!}
{(m-k+u)!\,(k+u+1)!\,(k-u)!} \\
&=&
(-1)^{p+u+1}2^{p+1/2}\sqrt{m!\,(m+2u+1)!}\sum_{l=0}^{m}\frac{(-1)^l(l+u+p+1)!}
{(m-l)!\,(l+2u+1)!\,l!} \\
&=&
(-1)^{p+u+1}2^{p+1/2}(p+u+1)!\sqrt{\frac{m!}{(m+2u+1)!}}{{m+u-p-1}\choose{m}}\\
&=&
(-1)^{p+u+m+1}2^{p+1/2}(p+u+1)!\sqrt{\frac{m!}{(m+2u+1)!}}
{{p-u}\choose{m}}\\
&=&0,\;\;\;\text{ if and only if $p\ge u$ and $m>p-u$.}
\end{eqnarray*}
Now we know exactly the matrix elements of any $Y^{(j)}$.

\

We will now proceed to the proof of the formula of Cahill and Glauber, namely the relation
\begin{equation}\label{matrixelements}
\langle n|D(z)K^\lambda D(z)^*|m\rangle = \sqrt{\frac{n!}{m!}}e^{-(1-\lambda)|z|^2}(1-\lambda)^{m-n+1}\cc{z}^{m-n}\lambda^nL_n^{m-n}(-(1-\lambda)^2|z|^2/\lambda),
\end{equation}
where $n,m\in\N$. For any $\lambda\in \C$ define  
$$
K^\lambda:=(1-\lambda)\lambda^N=(1-\lambda)\sum_{k=0}^\infty \lambda^n |n\rangle \langle n|,
$$
on the domain
$$
D(K^\lambda)= \bigg\{\vp\in \hil \bigg| \sum_{n=0}^\infty |\lambda|^{2n} |\langle n|\vp\rangle|^2<\infty \bigg\}.
$$
It follows from the usual spectral theory that $(K^\lambda)^*=K^{\cc{\lambda}}$.
In addition,
\begin{equation}\label{domain}
D(z)|k\rangle\in D(K^\lambda), \ \ \ k\in \N, \, z\in \C.
\end{equation}
Indeed, we have
$$
|\langle n|D(z)|k\rangle|^2=\frac{k!}{n!}e^{-|z|^2}|z|^{2(n-k)}|L_k^{n-k}(|z|^2)|^2,  \  \ n\geq k,
$$
and $|L_k^{n-k}(|z|^2)|^2\leq \left(\frac{n!}{k!(n-k)!}\right)^2e^{|z|^2}$ (see e.g. \cite[p. 786, 22.14.13]{Abramowitz}), so that
$$
\sum_{n=k}^\infty |\lambda|^{2n} |\langle n|D(z)|k\rangle|^2
\leq \frac{1}{k!}\sum_{n=k}^\infty |\lambda|^{2n}\frac{n!}{((n-k)!)^2}|z|^{2(n-k)}
= \frac{|\lambda|^{2k}}{k!}\sum_{n=0}^\infty |\lambda z|^{2n}\frac{(n+k)!}{(n!)^2}<\infty,
$$
the last series converging by the ratio test. Because of \eqref{domain}, the operator
$D(z)K^\lambda D(z)^*$ is densely defined, with the domain containing all the number states.
Hence, the left hand side of the relation \eqref{matrixelements} makes sense.

By \eqref{domain}, also the coherent states belong to the domain of $D(z)K^\lambda D(z)^*$. Using \eqref{domain} again, we can write
\begin{eqnarray}
\langle \beta |D(z)K^\lambda D(z)^*|\beta\rangle &=&
 \sum_{m,n=0}^\infty \langle \beta|n\rangle \langle n|D(z)K^\lambda D(z)^*|m\rangle \langle m|\beta\rangle\nonumber\\
&=& e^{-|\beta|^2}\sum_{m,n=0}^\infty \frac{\beta^m\cc{\beta}^n}{\sqrt{m!n!}}\langle n|D(z)K^\lambda D(z)^*|m\rangle\label{series}.
\end{eqnarray}

To prove \eqref{matrixelements}, one first calculates
\begin{eqnarray*}
&&\langle \beta |D(z)K^\lambda D(z)^*|\beta\rangle= \langle \beta-z|K^\lambda|\beta-z\rangle
= (1-\la)\sum_{n=0}^\infty \lambda^n \langle \beta-z|n\rangle \langle n|\beta-z\rangle\\
&=& (1-\la)e^{-\frac 12 (|\beta-z|^2+|\beta-z|^2)}\sum_{n=0}^\infty \lambda^n \frac{|\beta-z|^{2n}}{n!} \\
&=& (1-\la)\exp[(\lambda-1)|\beta-z|^2] = \exp[(\lambda-1)(|\beta|^2+|z|^2-\cc{\beta}z-\cc{z}\beta)]\\
&=& (1-\la)\exp[-|\beta|^2+(\lambda-1)|z|^2]\exp[\beta((1-\lambda)\cc{z}+\lambda\cc{\beta})]\exp[(1-\lambda)z\cc{\beta}]\\
&=& (1-\la)\exp[-|\beta|^2+(\lambda-1)|z|^2]\sum_{m=0}^\infty \frac{\beta^m}{m!}[(1-\lambda)\cc{z}+\lambda\cc{\beta}]^m \exp[(1-\lambda)z\cc{\beta}]\\
&=& (1-\la)e^{-|\beta|^2+(\lambda-1)|z|^2}\sum_{m=0}^\infty \frac{\beta^m}{m!}[(1-\lambda)\cc{z}]^m\left(1+\frac{\lambda\cc{\beta}}{(1-\lambda)\cc{z}}\right)^m \exp\left[\left(\frac{((1-\lambda)|z|)^2}{\lambda}\right)\frac{\cc{\beta}\lambda}{(1-\lambda)\cc{z}}\right],
\end{eqnarray*}
where one has to assume that $z\neq 0$, $\lambda\neq 0$, and $\lambda\neq 1$.
Now one can use the formula $(1+y)^m \exp[-xy] = \sum_{n=0}^\infty L_n^{m-n}(x)y^n$, which holds for
$|y|<1$ \cite[p.1038, 8.975(2)]{Gradshteyn}, with $x=-\frac{((1-\lambda)|z|)^2}{\lambda}$ and $y=\frac{\lambda\cc{\beta}}{(1-\lambda)\cc{z}}$. This then requires that $|\beta|<|(\lambda^{-1}-1)z|$. One obtains
\begin{eqnarray*}
&&\langle \beta |D(z)K^\lambda D(z)^*|\beta\rangle\\
&=&(1-\la) e^{-|\beta|^2+(\lambda-1)|z|^2}\sum_{m,n=0}^\infty \frac{\beta^m\cc{\beta}^n}{m!}[(1-\lambda)\cc{z}]^m\left(\frac{\lambda}{(1-\lambda)\cc{z}}\right)^n L_n^{m-n}\left(\frac{((1-\lambda)|z|)^2}{-\lambda}\right)\\
&=&(1-\la)e^{-|\beta|^2}\sum_{m,n=0}^\infty \frac{\beta^m\cc{\beta}^n}{\sqrt{m!n!}}\left(\sqrt{\frac{n!}{m!}}e^{(\lambda-1)|z|^2}[(1-\lambda)\cc{z}]^{m-n}\lambda^n L_n^{m-n}\left(-\frac{(1-\lambda)^2|z|^2}{\lambda}\right)\right).\\
\end{eqnarray*}
Putting $z=re^{i\theta}$, we get
\begin{eqnarray*}
&&\langle \beta |D(z)K^\lambda D(z)^*|\beta\rangle\\
&=&e^{-|\beta|^2}\sum_{m,n=0}^\infty \frac{\beta^m\cc{\beta}^n}{\sqrt{m!n!}}\left(e^{i(n-m)}(1-\la)\sqrt{\frac{n!}{m!}}e^{(\lambda-1)r^2}[(1-\lambda)r]^{m-n}\lambda^n L_n^{m-n}\left(-\frac{(1-\lambda)^2r^2}{\lambda}\right)\right).\\
\end{eqnarray*}
By comparing this with \eqref{series}, we get the result.


\begin{thebibliography}{99}

\bibitem{Abramowitz} M. Abramowitz, I. A. Stegun (eds.), {\em Handbook of Mathematical Functions}, National Bureau of Standards, Applied Mathematics
Series - 55, Tenth printing with corrections, 1972.
\bibitem{Albini} P. Albini, E. de Vito, A. Toigo, Quantum homodyne tomography as an informationally complete positive-operator-valued measure, {\em J. Phys. A: Math. Theor.} {\bf 42} (2009) 295302.
\bibitem{Armstrong} B. H. Armstrong, Spectrum line profiles: the Voigt function, {\em J. Quant. Spectrosc. Radiat. Transfer} {\bf 7} (1967) 61-88.
\bibitem{Bertrand1987} J. Bertrand, P. Bertrand, A tomographic approach to Wigner's function, {\em Found. Phys} {\bf 17} (1987) 397-405.
\bibitem{Birman} M. S. Birman, M. Z. Solomjak, {\em Spectral Theory of Self-Adjoint Operators in Hilbert Space},
D. Reidel, Dordrecht, 1987.
\bibitem{Busch} P. Busch, P. Lahti, The determination of the past and the future of a physical system in quantum mechanics, {\em Found. Phys.} {\bf 19} (1989) 633-678. 
\bibitem{Cassinelli} G. Cassinelli, G.M. D'Ariano, E. De Vito, A. Levrero,
Group theoretical quantum tomography, {\em J. Math. Phys.} {\bf 41} (2000) 7940-7951.
\bibitem{Cahill1969} K. E. Cahill, R. J. Glauber, Ordered expansions in boson amplitude operators, {\em Phys. Rev.} {\bf 177} (1969) 1857-1881.
\bibitem{Cooke}R. G. Cooke, {\em Infinite Matrices and Sequence Spaces}, Dover Publications, Inc., New York, 1955.
\bibitem{Glauber1969} K. E. Cahill, R. J. Glauber, Density operators and quasiprobability distributions, {\em Phys. Rev.} {\bf 177} (1969) 1882-1902.
\bibitem{Dariano1994} G. M. D'Ariano, C. Macchiavello, M. G. A. Paris, Detection of the density matrix through optical homodyne tomography without filtered back projection, {\em Phys. Rev. A} {\bf 50} (1994) 4298-4302.
\bibitem{Macchiavello1995} G. M. D'Ariano, C. Macchiavello, M. G. A. Paris, Optimized phase detection, {\em Phys. Lett. A} {\bf 198} (1995) 286-294.
\bibitem{DaMa1996} G. M. D'Ariano, S. Mancini, P. Tombesi, V. I. Man'ko, Reconstructing the density operator by using generalized field quadratures,  {\em Quant. Semiclass. Opt.} {\bf 8} (1996) 1017-1027.
\bibitem{Duran} A. J. Duran, A bound on the Laguerre polynomials, {\em Studia Math.} {\bf 100} (1991) 169-181.
\bibitem{Gradshteyn} I. S. Gradshteyn, I. M. Ryzhnik, {\em Table of Integrals, Series, and Products},
Corrected and Enlarged Edition, Academic Press, Inc., Orlando, 1980.
\bibitem{KL2008} J. Kiukas, P. Lahti, On the moment limit of quantum observables, with an application to the balanced homodyne detection, {\em J. Mod. Opt.} {\bf 55} (2008) 1175-1198.
\bibitem{eightport} J. Kiukas, P. Lahti, A note on the measurement of phase space observables with an eight-port homodyne detector, {\em J. Mod. Opt.} {\bf 55} (2008) 1891-1898.
\bibitem{KLP2008} J. Kiukas, P. Lahti, J.-P. Pellonp\"a\"a, A proof for the informational completeness of the rotated quadrature observables, 
{\em J. Phys. A: Math. Theor.} {\bf 41} (2008) 175260.
\bibitem{Kuhn1994} H. K\"uhn, D.-G. Welsch, W. Vogel, Determination of density matrices from field distributions and quasiprobabilities, {\em J. Mod. Opt.} {\bf 41} (1994) 1607-1613.
\bibitem{Lahti} P. Lahti, J.-P. Pellonp\"a\"a, Continuous variable tomographic measurements, {\em Phys. Lett. A} {\bf 373} (2009) 3435-3438. 
\bibitem{Leibfried1996}D. Leibfried, D. M. Meekhof, B. E. King, C. Monroe, W. M. Itano, D. J. Wineland,
Experimental determination of the motional quantum state of a trapped atom,
{\em Phys. Rev. Lett.} {\bf 77} (1996) 4281-4285.
\bibitem{Leonhardt1996} U. Leonhardt, M. Munroe, T. Kiss, Th. Richter, M. G. Raymer, Sampling of photon statistics and density matrix using homodyne detection, {\em Opt. Comm.} {\bf 127} (1996) 144-160.
\bibitem{Leonhardt1993} U. Leonhardt, H. Paul, Realistic optical homodyne measurements and quasiprobability distibutions, {\em Phys. Rev. A} {\bf 48} (1993) 4598-4604.
\bibitem{Leonhardt} U. Leonhardt, {\em Measuring the Quantum State of Light}, Cambridge University Press, 1997.
\bibitem{DAriano} U. Leonhardt, H. Paul, G. M. D'Ariano, Tomographic
reconstruction of the density matrix via pattern functions, {\em Phys. Rev. A}
{\bf 52} (1995) 4899-4907.
\bibitem{Lougovski2003} P. Lougovski, E. Solano, Z. M. Zhang, H. Walther, H. Mack, W. P. Schleich, Fresnel representation of the Wigner function: an operational approach, {\em Phys. Rev. Lett.} {\bf 91} (2003) 010401. 
\bibitem{Mancini1996} S. Mancini, V. I. Man'ko, P. Tombesi, Symplectic tomography as classical approach to quantum systems, {\em Phys. Lett. A} {\bf 213} (1996) 1-6.
\bibitem{Mancini1997} S. Mancini, P. Tombesi, V. I. Man'ko, Density matrix from photon number tomography, {\em Europhys. Lett.} {\bf 37} (1997) 79-83.
\bibitem{McCabe} J. H. McCabe, A continued fraction expansion, with a truncated error estimate, for Dawson's integral, {\em Math. Comp.} {\bf 28} (1974) 811-816.
\bibitem{Milone} L. A. Milone, A. A. E. Milone, Evaluation of Dawson's function, {\em Astrophys. Space Sci.} (1988) 189-191.
\bibitem{DeNicola} S. De Nicola, R. Fedele, M. A. Man'ko, V. I. Man'ko, Fresnel tomography: a novel approach to the wave function reconstruction based on Fresnel representation of tomograms, arXiv:0503043.
\bibitem{Paris} M. G. A. Paris, Quantum state measurement by realistic heterodyne detection, {\em Phys. Rev. A} {\bf 53} (1996) 2658-2663.
\bibitem{Pellonpaa} J.-P. Pellonp\"a\"a, Quantum tomography, phase space observables, and generalized Markov kernels, {\em J. Phys. A: Math. Theor.}, in press, arXiv:0906.2101.
\bibitem{Prugovecki} E. Prugove\v cki, Information-theoretical aspects of quantum measurement, {\em Int. J. Theor. Phys.} {\bf 16} (1977) 321-331.
\bibitem{Putnam} C. R. Putnam, {\em Commutation Properties of Hilbert Space Operators and Related Topics}, Springer-Verlag, Berlin, 1967.
\bibitem{Riordan} J. Riordan, {\em Combinatorial Identities}, John Wiley \& Sons, inc., New York, 1968.
\bibitem{Schreier} F. Schreier, The Voigt function and complex error function: a comparison of computational methods, {\em J. Quant. Spectrosc. Radiat. Transfer.} {\bf 48} (1992) 743-762.
\bibitem{Smithey1993} D.T. Smithey, M. Beck, M.G.Raymer, A. Faridina, Measurement of the Wigner distribution and the density 
matrix of a light mode using optical homodyne tomography: application to squeezed states and the vacuum,
{\em Phys. Rev. Lett.} {\bf 70} (1993) 1244-1247.
\bibitem{Atlas} J. Spanier, K. B. Oldham, {\em An Atlas of Functions}, Hemisphere Publishing Company, 1987.
\bibitem{VR1989} K. Vogel, H. Risken, Determination of quasiprobability distributions in terms of probability distributions for
the rotated quadrature phase, {\em Phys. Rev. A} {\bf 40} (1989) 2847-2849.
\bibitem{Werner} R. Werner, Quantum harmonic analysis on phase space, {\em J. Math. Phys.} {\bf 25} (1984) 1404-1411.
\bibitem{Welsch} D.-G. Welsch, W. Vogel, T. Opatrn\'y, Homodyne detection and quantum state reconstruction, arXiv:0907.1353.
\end{thebibliography}
\end{document}